\documentclass[twoside,english,aps,manuscript,superscriptaddress,preprint,notitlepage,11pt]{revtex4-1}
\usepackage[T1]{fontenc}
\usepackage{mathrsfs}
\usepackage{amsmath}
\usepackage{amssymb}
\usepackage{graphicx}
\usepackage{esint}

\makeatletter
\@ifundefined{textcolor}{}
{%
 \definecolor{BLACK}{gray}{0}
 \definecolor{WHITE}{gray}{1}
 \definecolor{RED}{rgb}{1,0,0}
 \definecolor{GREEN}{rgb}{0,1,0}
 \definecolor{BLUE}{rgb}{0,0,1}
 \definecolor{CYAN}{cmyk}{1,0,0,0}
 \definecolor{MAGENTA}{cmyk}{0,1,0,0}
 \definecolor{YELLOW}{cmyk}{0,0,1,0}
}

\usepackage[hypertexnames=false]{hyperref}
\hypersetup{colorlinks=true,citecolor=blue,linkcolor=blue,urlcolor=blue} 

\AtBeginDocument{
\renewcommand{\ref}[1]{\autoref{#1}}

}
\makeatother

\usepackage{babel}
\begin{document}

\title{Tunable magnetization relaxation of Fe$_{2}$Cr$_{1-x}$Co$_{x}$Si
half-metallic Heusler alloys by band structure engineering \smallskip{}
}

\author{Shikun He}

\thanks{These authors contributed equally}

\affiliation{Division of Physics and Applied Physics, School of Physical and Mathematical
Sciences, Nanyang Technological University, Singapore 637371}

\affiliation{Data Storage Institute, Agency for Science, Technology and Research
(A{*}STAR), 2 Fusionopolis Way 08-01 Innovis, Singapore 138634}

\author{Yifan Liu}

\thanks{These authors contributed equally}

\affiliation{Department of Electrical and Computer Engineering, National University
of Singapore, Singapore 117582}

\author{Yuhong Zheng }

\affiliation{Department of Electrical and Computer Engineering, National University
of Singapore, Singapore 117582}

\author{Qing Qin}

\affiliation{Department of Materials Science and Engineering, National University
of Singapore, Singapore 117574}

\author{Zhenchao Wen}

\affiliation{Center for Spintronics Research Network (CSRN), Tohoku University,
Sendai 980-8577, Japan.}

\author{Qingyun Wu}

\affiliation{Department of Physics, National University of Singapore, Singapore
117542}

\author{Yi Yang}

\affiliation{Data Storage Institute, Agency for Science, Technology and Research
(A{*}STAR), 2 Fusionopolis Way 08-01 Innovis, Singapore 138634}

\author{Yupu Wang}

\affiliation{Department of Electrical and Computer Engineering, National University
of Singapore, Singapore 117582}

\author{YuanPing Feng}

\affiliation{Department of Physics, National University of Singapore, Singapore
117542}

\author{Kie Leong Teo}

\affiliation{Department of Electrical and Computer Engineering, National University
of Singapore, Singapore 117582}

\email{eleteokl@nus.edu.sg}

\selectlanguage{english}%

\author{Christos Panagopoulos}

\affiliation{Division of Physics and Applied Physics, School of Physical and Mathematical
Sciences, Nanyang Technological University, Singapore 637371}

\email{christos@ntu.edu.sg}

\selectlanguage{english}%
\begin{abstract}
We report a systematic investigation on the magnetization relaxation
properties of iron-based half-metallic Heusler alloy Fe$_{2}$Cr$_{\textrm{1-x}}$Co$\mbox{\ensuremath{_{x}}}$Si
(FCCS) thin films using broadband angular-resolved ferromagnetic resonance.
Band structure engineering through Co doping ($x$) demonstrated by
first principles calculations is shown to tune the intrinsic magnetic
damping over an order of magnitude namely, $1\times10^{-2}${\normalsize{}-$8\times10^{-4}$}.
Notably, the intrinsic damping constants for samples with high Co
concentration are amongst the lowest reported for Heusler alloys and
even comparable to magnetic insulator yttrium iron garnet. Furthermore,
a significant reduction of both isotropic and anisotropic contributions
of extrinsic damping of the FCCS alloys was found in the FCCS films
with x = 0.5 \textasciitilde{}0.75, which is of particular importance
for applications. These results demonstrate a practical recipe to
tailor functional magnetization for Heusler alloy-based spintronics
at room temperature.
\end{abstract}
\maketitle

\section{introduction.}

Control of magnetization and its dynamics in nanomagnets by injecting
an electrical current is increasingly important \citep{bergerPRB,Slonczewski_STT_1996}.
Operations based on spin transfer torque (STT) are energy efficient
with superior scalability for high density device applications compared
to conventional methods employing magnetic fields \citep{Ikeda_2010_Namat,Review_MRAM_wang2013low,woledge_2011,NaNano_comment_kent2015new}.
A promising application of STT is non-volatile magnetic random access
memory (MRAM) \citep{20nmMRAM_gajek2012spin,thomas2014perpendicular}.
Key parameter here is magnetic damping; it determines the critical
switching current and speed of magnetization reversal \citep{Slonczewski_STT_1996,Mangin_STT_TMR}.
Hence, a magnetic material with tunable low damping is needed. Magnetic
relaxation is primarily due to spin-orbit coupling (SOC) however,
materials and device guidance from theoretical calculations is hindered
by the complicity of inter- and intraband scattering, defects, disorder,
interface effects, and geometrical confinement \citep{Theory_PRL_Starikov,damping_breathMode_PRL}.
A practical criterion for engineering magnetic damping is $\alpha\propto\xi^{2}D(E_{F})$
\citep{kambersky1984fmr,Review_JphyD_2010}, where $\xi$ is the SOC
parameter and $D(E_{F})$ the density of states (DOS) at the Fermi
level. Although Fe, Co and Ni render the window for tunable SOC in
3d-based materials formidably narrow, there have been encouraging
developments through the correlation between damping and electronic
structure, as demonstrated in FeV and CoFe binary alloys \citep{PRL_FeV_2007,Justin_NatPhy}.
Still, reducing and tuning magnetic damping in metallic materials
can be difficult due to magnon scattering by high density conduction
electrons \citep{Theory_PRL_Starikov}. Notably, the lowest reported
damping for CoFeB, the material employed by industry for MRAM development,
is approximately 0.004 \citep{NTU_FMR,okada2017magnetization}.

Half-metallic Heusler alloy is a semiconductor for one spin projection
and a metal for the other \citep{RMP_katsnelson2008half} hence, the
band structure is expected to result in high spin polarization and
low damping \citep{Calculation_lowdamping_halfMetal,jourdan2014direct_HM,Wen_PRA,WenAdvanceMaterial,qin2017ultra}.
Both properties are favorable for STT \citep{Mangin_STT_TMR}. However,
except for a few Co-based Heusler alloys, the damping constant is
larger than for Fe films (0.002), probably due to band broadening
by structural defects, chemical disorder and magnon excitation \citep{Review_JphyD_2010,CFA_PRB_TMS_Cos,CFA_lowdamping_2013,CFA_DOS,CFA_2016_damping,kohler2016tunable}.
Furthermore, Heusler alloys usually exhibit large anisotropy in the
ferromagnetic resonance (FMR) linewidth \citep{CFA_DOS,CoFeSi_anisotropy_2014},
which may lead to large variation in performance among devices. Therefore,
tuning and reducing intrinsic damping, and a minimum extrinsic magnetic
relaxation are required for reliable operations. Fe-based Heusler
alloys are expected to possess half-metallic band structure \citep{luo2007DOS}.
Also, these materials hold great promise for high interfacial perpendicular
anisotropy \citep{wang2014perpendicular}, an indispensable property
for next generation STT-MRAM, at a Fe-MgO interface \citep{yang2011first,koo2013large}.
However, a comprehensive magnetization relaxation study is still lacking
for Fe-based Heusler alloys.

Here, we tune the band structure of Fe$_{2}$Cr$_{1-x}$Co$_{x}$Si
(FCCS) by Co doping ($x$). Dynamic and static properties were investigated
using angle-resolved broadband FMR. We show ordered states ($B2$
and $L2_{1}$) exhibit considerably lower damping compared to disordered
states ($A2$). Crucially, Co doping allows control of the intrinsic
damping constant by an order of magnitude ($1\times10^{-2}-8\times10^{-4}$).
Furthermore, both isotropic and anisotropic extrinsic contributions
to the in-plane magnetization relaxation of quaternary FCCS can be
significantly reduced for $x$ between 0.5 and 0.75.

\section{Experiment }

The samples were prepared in an ultra-high vacuum magnetron sputtering
system with base pressure $10^{-7}$ Pa. Prior to deposition, the
MgO (100) substrate was heated at 600$^{\circ}C$ for one hour. Subsequently,
a 40nm Cr buffer layer was deposited followed by one hour in-situ
annealing at 700$^{\circ}C$ . The quaternary Heusler alloy FCCS (30$\,$nm)
layer with variable Co doping, was deposited by co-sputtering Fe$_{2}$CrSi
and Fe$_{2}$CoSi. The deposition speed of individual target, hence
the nominal value of $x$, was tuned by adjusting the sputtering power.
Total deposition rate was approximately 0.3$\,$$\textrm{Å}$/s
for smooth growth. An in-situ annealing procedure was performed at
$T_{\textrm{ia}}$ for 30 minutes to develop ordered $B2$ or $L2_{1}$
structures. A series of $T_{\textrm{ia}}$ (350-500$^{\circ}C$) was
used for $x$=0.5 whereas $T_{\textrm{ia}}$ was fixed at 450$^{\circ}C$
for all other samples. Finally, the FCCS layer was capped by Ru (3$\,$nm)
to avoid oxidation. 

Our homebuilt angular-resolved FMR system equipped with an Agilent
E8361C vector network analyzer (VNA) can measure with frequency up
to 40$\,$GHz. We use grounded coplanar waveguide with a nominal impedance
of 50$\,\Omega$ and a center conductor of width 100$\,$$\mu$m \citep{NTU_FMR}.
A customized sample holder is attached to a motorized stage for out-of-plane
field rotation. In addition, an in-plane sample manipulator is used
to rotate and mount the sample face down on the waveguide automatically.
The accuracy of in-plane and out-of-plane field orientations is better
than 0.1 degree. Using angle-resolved broadband FMR, we determine
the intrinsic damping by measuring with magnetization perpendicular
to the film plane, whereas the anisotropy in magnetization relaxation
was investigated by varying the magnetization orientation within the
film plane.

\section{Results and discussion}

\subsection{Structure and chemical ordering}

\begin{figure*}
\centering{}\includegraphics[width=0.85\columnwidth]{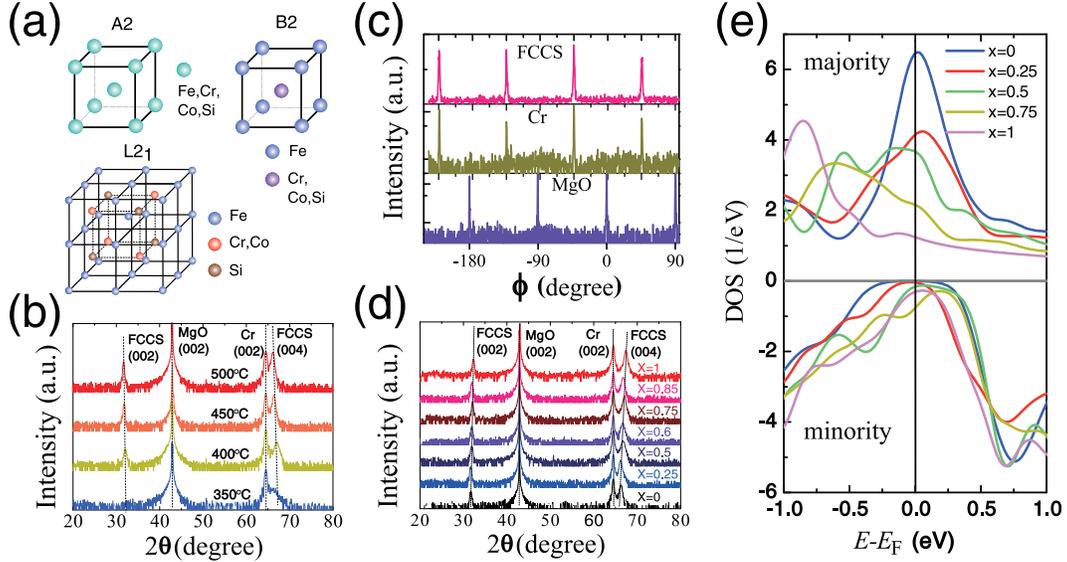}\protect\caption{Structures of the films with stacks of MgO(001)-substrate (30nm)/Cr
(40nm)/Fe$_{2}$Cr$_{1-x}$Co$_{x}$Si (30nm)/Ru (3nm).\textbf{ }(a)
Schematics of the disordered $A2$, partially ordered $B2$ and ordered
$L2_{1}$ structures. (b) XRD $\theta-2\theta$ scans for Fe$_{2}$Cr$_{0.5}$Co$_{0.5}$Si
thin films with various in-situ annealing temperatures. (c) $\phi$
scans of FCCS, Cr and MgO of MgO(001)-substrate/Cr(40nm)/ Fe2Cr0.5Co0.5Si(30nm)/
Ru(3nm) for the (111) plane. The in-situ annealing temperature is
450$^{\circ}C$. (d) XRD $\theta-2\theta$ scans for Fe$_{2}$Cr$_{1-x}$Co$_{x}$Si
with various Co concentrations. The films were in-situ annealed at
450$^{\circ}C$. (e) Density of states (DOS) of FCCS alloys determined
by first principle calculations. \label{fig:XRD}}
\end{figure*}

As shown in \ref{fig:XRD}(a), there are three structures in a FCCS
Heusler alloy film according to their atomic ordering: fully ordered
full-Heulse $L2_{1}$ structure, partially disordered $B2$ and fully
disordered $A2$ structures.

FCCS(002) peaks of XRD $\theta-2\theta$ scans of Fe$_{2}$Cr$_{0.5}$Co$_{0.5}$Si
films (\ref{fig:XRD}(b)) indicate that ordering of the FCCS film
evolves from $A2$ to $B2$ or $L2_{1}$ for $T_{\textrm{ia}}$=400$^{\circ}C$.
The Cr and FCCS layers exhibit epitaxial growth at 45$^{\circ}$ with
respect to MgO substrate (XRD phi-scan in \ref{fig:XRD}(c) and See
Supplemental Material Fig. S4 for Transmission electron microscopy
results). The degree of $B2$ ordering ($S_{B2}$) and $L2_{1}$ ordering
($S_{L2_{1}}$) was calculated using the Webster model \citep{webster1971magnetic}.
Best crystalline structure with $S_{B2}$=0.95 and $S_{L2_{1}}$=0.87
was obtained for $T_{\textrm{ia}}$=450 (See Supplemental Material
Fig. S1(a) for $T_{\textrm{ia}}$ dependence of ordering ). Correspondingly,
the saturation magnetization $M_{s}$ increases with $T_{\textrm{ia}}$
reaching maximum at $T_{\textrm{ia}}$=450$^{\circ}C$ (See Supplemental
Material Fig. S2 for $T_{\textrm{ia}}$dependence of $M_{s}$). Furthermore,
$B2$ and $L2_{1}$ phases were obtained for all Co concentrations
with $T_{\textrm{ia}}$=450$^{\circ}C$ as inferred from \ref{fig:XRD}(d).
All samples exhibited high degree $B2$ ordering ($S_{B2}$>0.74)
and moderate $S_{L2_{1}}$ ordering (See Supplemental Material Fig.
S1(b) for Co concentration dependence of ordering). B2 phase characterized
by Y-Z atomic disorder in Fe-based X$_{2}$YZ full Heusler alloy is
also favorable for device applications since it smears the DOS spectral
shapes, does not degrade half-metallicity significantly and has little
influence on damping \citep{disorderEffect,sakuma2015first,luo2007DOS}.
Our first principles calculations (See Supplemental Material S3 for
calculation methods) have shown that the band structures of $L2_{1}$
FCCS were sucessfully tuned by Co concentration. As can be clearly
seen from \ref{fig:XRD}(e), Fe$_{2}$CrSi and Fe$_{2}$Cr$_{0.75}$Co$_{0.25}$Si
exhibit half-metallicity whereas for $x\geqslant0.5$, the minority
band DOS at the Fermi level is non-zero and increases with $x$. On
the other hand, the total DOS at Fermi level dominated by the majority
band decreases monotonically with $x$.

\subsection{Dynamic properties of Fe$_{2}$Cr$_{0.5}$Co$_{0.5}$Si. }

\begin{figure}
\centering{}\includegraphics[width=0.85\columnwidth]{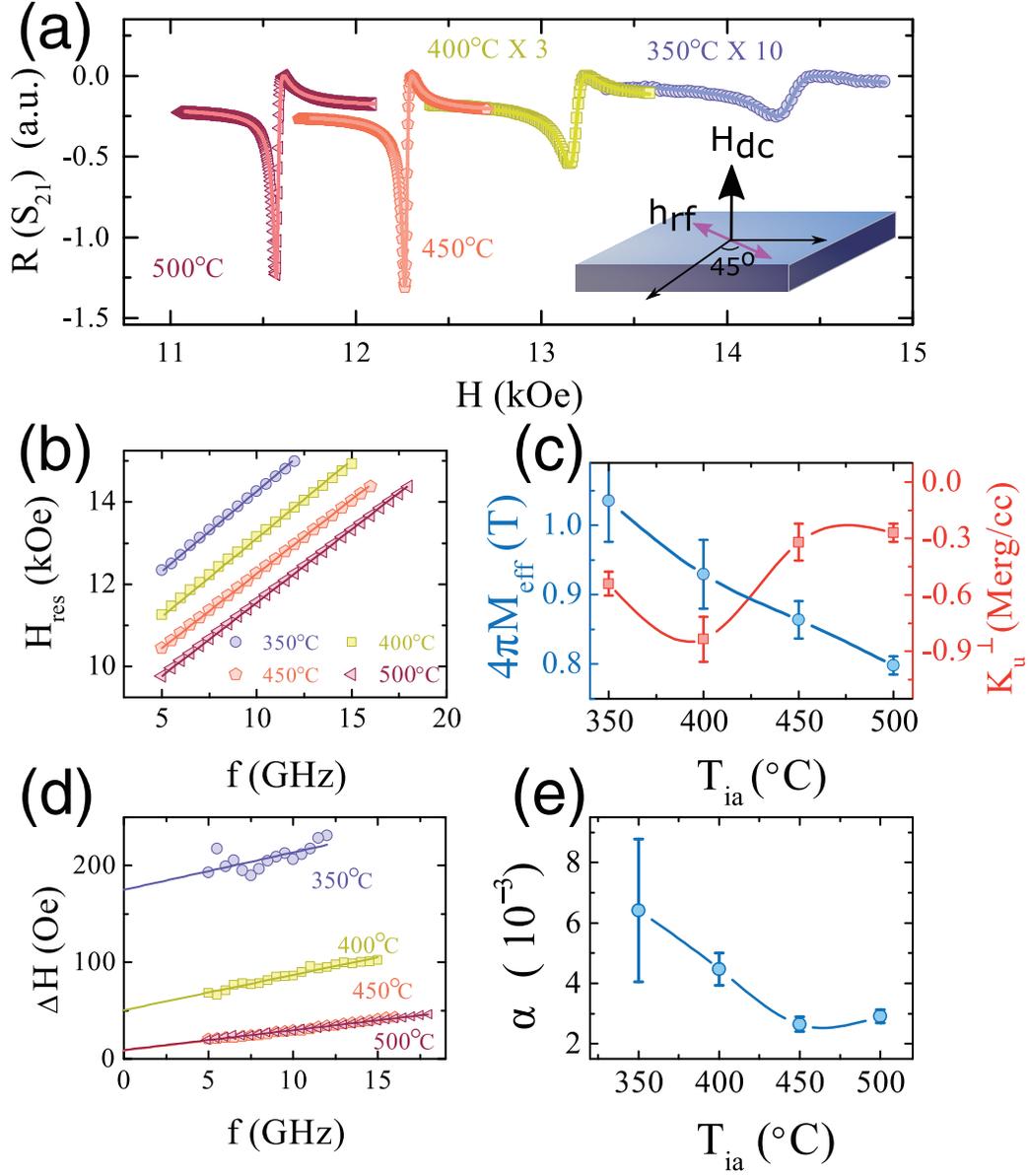}\protect\caption{Room temperature out-of-plane FMR results of Fe$_{2}$Cr$_{0.5}$Co$_{0.5}$Si
for different in-situ annealing temperature $T{}_{\textrm{ia}}$.
(a) Field dependence of the amplitude of FMR spectra for $f=10$ GHz.
The inset is an illustration of the measurement configuration. (b)
Frequency dependence of FMR resonance field. (c) Effective magnetization
and perpendicular anisotropy versus $T_{\mbox{ia}}$. (d) Frequency
dependence of FMR linewidth.(e) Damping as a function of $T_{\mbox{ia}}$
\label{fig:FMR-Tia}}
\end{figure}

Room temperature out-of-plane FMR spectra were taken with $\varphi_{H}=\pi/4$.
\ref{fig:FMR-Tia}(a) shows FMR data for Fe$_{2}$Cr$_{0.5}$Co$_{0.5}$Si
at different $T_{\textrm{ia}}$. Stronger signals and narrower peaks
are observed for $T_{\textrm{ia}}$ > 400$^{\circ}C$. The narrowest
spectrum is for $T_{\textrm{ia}}$ = 450$^{\circ}C$ , in correspondence
with the structure and magnetization measurements \citep{YuPu_wang2014structural}.
The resonance field $H_{\textrm{Res}}$ and full width at half maximum
(FWHM) $\triangle H$ were extracted by modified Lorentz fit (See
Supplemental Material S2 for fitting method). The frequency dependences
of $H_{\textrm{Res}}$ are shown in \ref{fig:FMR-Tia}(b). The Kittel
equation describing the resonance condition for out-of-plane configuration
is given by 

\begin{equation}
2\pi f=\gamma(H-4\pi M_{{\rm eff}})\label{eq:Kittel}
\end{equation}
indicating the linear relation between resonance frequency and field
Here $\gamma{\rm =g}\cdot e/2m_{e}$ is the gyromagnetic ratio determined
by Lande g factor, electron charge $e$ and mass $m_{e}$. $4\pi M_{{\rm eff}}$
the effective saturation magnetization $4\pi M_{{\rm eff}}=4\pi M_{s}-2K_{u}^{\bot}/M_{s}=4\pi M_{s}-H_{k}$;
$K_{u}^{\bot}$ is the perpendicular anisotropy and $H_{k}$ the anisotropy
field. The data are well fitted by straight lines. Using the slope,
intercept of the fit, and $M_{s}$ which we measure using a vibrating
sample magnetometer (See Supplemental Material Fig. S2), we determined
the value of g, $4\pi M_{{\rm eff}}$ and $K_{u}^{\bot}$. As seen
from \ref{fig:FMR-Tia}(c), the films possess a negative $K_{u}^{\bot}$(easy
axis in the film plane). We find the corresponding $H_{k}$ to be
of the order of kOe, considerably larger than the in-plane anisotropy
field, which we discuss later. We believe $K_{u}^{\bot}$ originates
from interfacial stress and varies with $T_{\textrm{ia}}$ due to
the annealing temperature dependence of ordering and lattice constant
\citep{CFA_PRB_TMS_Cos}. \ref{fig:FMR-Tia}(d) indicates the FMR
linewidth increases linearly with frequency for all samples. This
allows the determination of damping constant $\alpha$ through: 

\begin{equation}
\Delta H_{{\rm OP}}=\frac{4\pi}{\gamma}\alpha f+\Delta H_{0}\label{eq:damping}
\end{equation}
Here, $\alpha$ is the coefficient of the viscous phenomenological
dissipation term after Gilbert\citep{Gilbert1955}. $\Delta H_{0}$
is the inhomogeneous broadening due to the dispersion of magnetic
properties. The damping constant for FCCS film with optimal $L2_{1}$
and $B2$ phase is $(2.6\pm0.3)\times10^{-3}$, i.e., 40 percent the
value of $A2$ phase ( $(6.5\pm0.2)\times10^{-3}$, \ref{fig:FMR-Tia}(e).
The results indicate the ordered half-metallic phase has lower damping,
suggesting atomic chemical disorder enhances damping \citep{Theory_PRL_Starikov}.
Clearly, annealing at $T_{{\rm ia}}\ge400^{\circ}C$ improves sample
uniformity as inferred from the $T_{\textrm{ia}}$ dependence of $\Delta H_{0}$. 

\subsection{Co doping dependence of magnetization relaxation of Fe$_{2}$Cr$_{1-x}$Co$_{x}$Si. }

\begin{figure}
\centering{}\includegraphics[width=1\columnwidth]{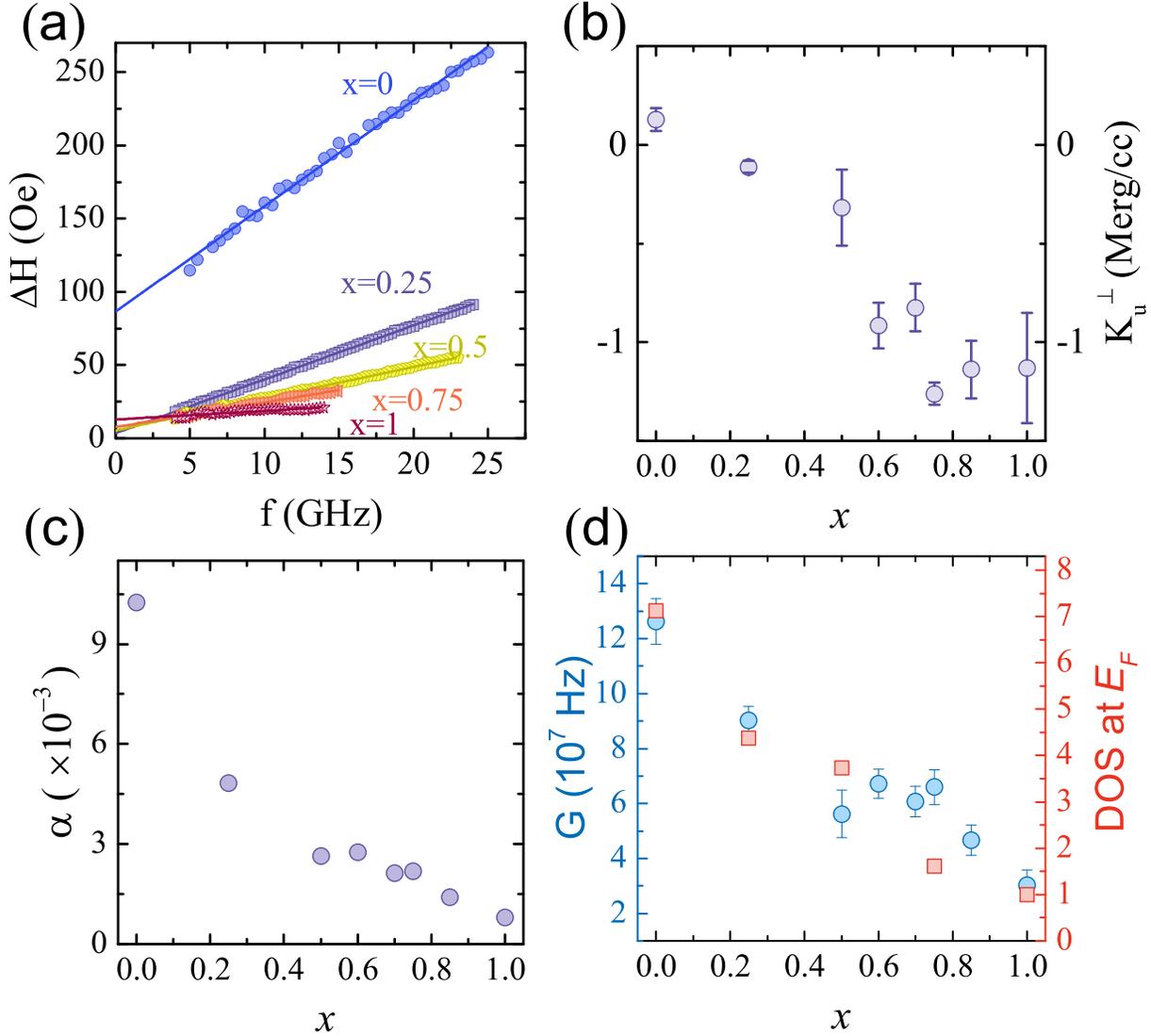}\protect\caption{Co concentration dependence of FMR results in Fe$_{2}$Cr$_{1-x}$Co$_{x}$Si.
(a) FMR linewidth versus frequency for various values of $x$. The
solid lines are linear fits to the data. (b) Perpendicular uniaxial
magnetic anisotropy versus $x$. (c) Damping constant $\alpha$ versus
$x$. (d) Relaxation frequency $G$ and total density of states at
the Fermi level versus $x$.\label{fig:FMR-Co-composition}}
\end{figure}

The out-of-plane FMR results with a magnetic field up to 2.1 T for
a series of FCCS samples are shown in \ref{fig:FMR-Co-composition}.
All samples were annealed at $T_{\textrm{ia}}$=450$^{\circ}C$. Again,
we find linear frequency dependences of the linewidth. The perpendicular
anisotropy constant and damping values shown in \ref{fig:FMR-Co-composition}(b)
and (c) were determined using the same procedure described earlier.
$K_{u}^{\bot}$ increases with $x$ may be attributed to the Co doping
dependence of tetragonal distortion \citep{TetraDistortionNi}, determined
by the reciprocal space mapping(See Supplemental Material Fig. S3
for structural informations). Damping for Fe$_{2}$CrSi ($x$=0) is
the largest $(1\pm0.03)\times10^{-2}$ . For $x$ =0.25, damping decreases
by approximately 50\% to $(4.8\pm0.1)\times10^{-3}$. Further increase
of Co doping results in a nearly linear reduction in damping. Fe$_{2}$CoSi
($x$ = 1) exhibits the lowest damping $(8\pm1)\times10^{-4}$; this
is considerably lower than 3d metals and remarkably close to the value
for high quality Y$_{3}$Fe$_{5}$O$_{12}$ (YIG) films \citep{heinrich_2011_YIG_4_cavity,YIG_WuMZ}. 

Relaxation frequency ${\rm G=}\gamma M_{s}\alpha$ is directly related
to the speed of relaxation and decreases by a factor of four with
increasing $x$. The lowest relaxation frequency (30$\,$MHz) observed
in Fe$_{2}$CoSi (\ref{fig:FMR-Co-composition}(d)) is comparable
to that in FeV films \citep{PRL_FeV_2007}. The offset to smaller
$\alpha$ and $G$ for $x$=0.5 is because $T_{\textrm{ia}}$ was
optimized at this particular value of $x$. Hence, damping can be
reduced further by fine tuning growth conditions.

Magnetic damping depends on SOC and DOS $G\propto\xi^{2}D(E_{F})$
\citep{kambersky1970landau}. Low $\xi$ in Heusler compounds because
of quenching of orbital moments \citep{Review_JphyD_2010} and suppression
of inter-band scattering \citep{damping_breathMode_PRL} due to relatively
high spin-polarization band structure may account for the overall
low damping. On the other hand, the $x$-dependence of relaxation
frequency is due to band structure engineering as shown in \ref{fig:XRD}
(d), indicated by the simultaneous decrease of relaxation frequency
and total DOS at the Fermi level with $x$. A small misalignment may
be due to band broadening by defects and chemical disorder. 

Similarly, according to Du et al., the Fermi level of FCCS ingots
moves from the bottom of conduction band to the top of valence band
with increasing $x$ for the minority band.\citep{du2013half} The
observed band structure tuning effect of Fe-based Heusler alloys also
agrees with earlier investigations on Co-based Heusler compounds.
\citep{kohler2016tunable}

\subsection{Anisotropic magnetization relaxation of Fe$_{2}$Cr$_{1-x}$Co$_{x}$Si. }

\begin{figure}
\centering{}\includegraphics[width=0.85\columnwidth]{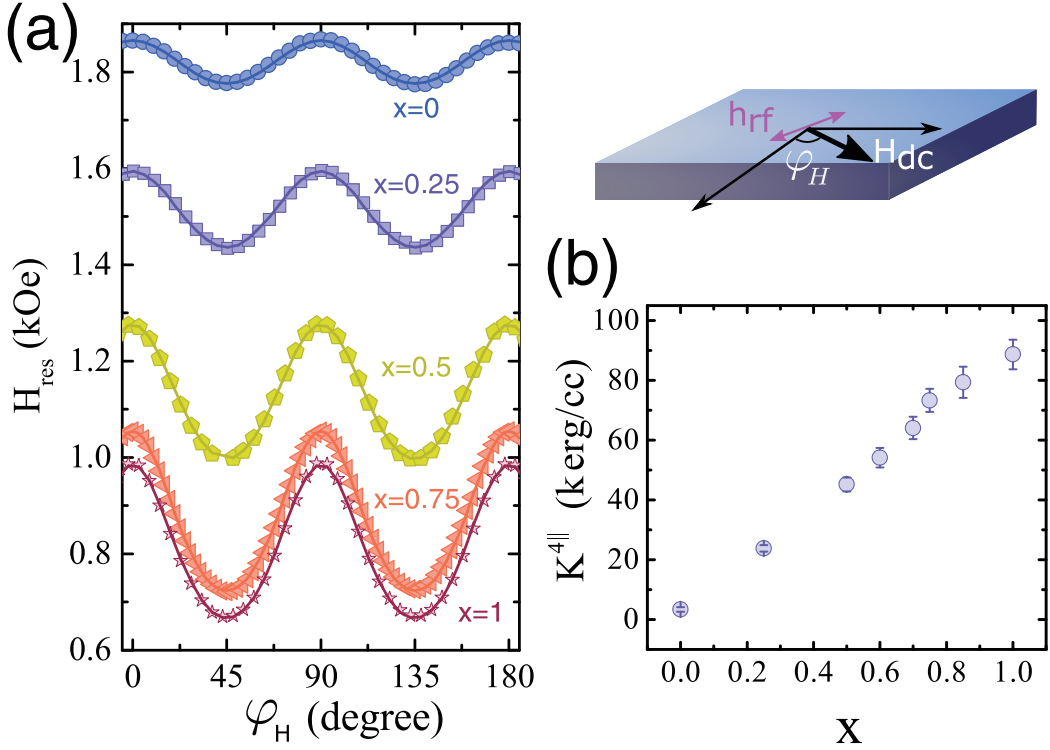}\protect\caption{Room temperature in-plane FMR resonance fields and anisotropic energy.
(a) In-plane FMR resonance field at $f$ =10 GHz as a function of
field direction for Fe$_{2}$Cr$_{1-x}$Co$_{x}$Si samples with different
Co concentration. Solid lines are calculated using \ref{eq:Kittel_angular}.
(b) Co concentration dependence of in-plane four-fold anisotropy energy.
The schematic on the top right shows the coordinates used in the measurements
and the orientations of DC and RF fields.\label{fig:IPR-ku}}
\end{figure}

\begin{figure*}
\centering{}\includegraphics[width=0.8\columnwidth]{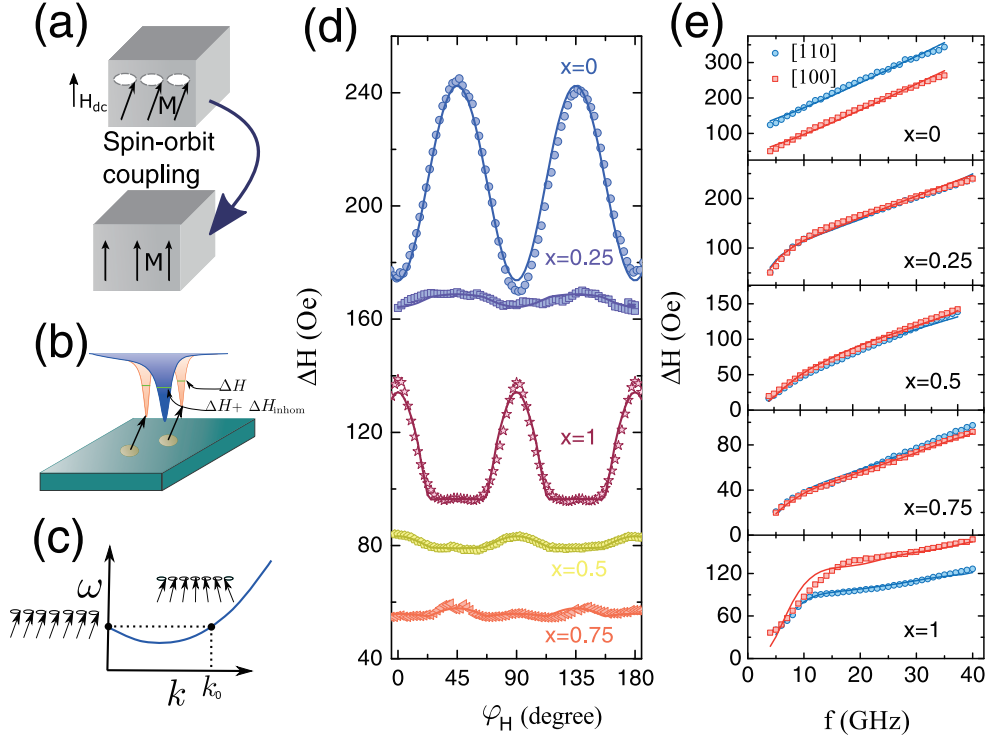}\protect\caption{In-plane angular and frequency dependence of FMR linewidth of Fe$_{2}$Cr$_{1-x}$Co$_{x}$Si.
(a-c) Illustrations of magnetization relaxation due to spin-orbit
coupling (a), inhomogeneity (b) and scattering by magnon excitation
(c). (a) Intrinsic damping is caused by spin-orbit coupling. (b) Sample
inhomogeneity broadens the FMR linewidth as the spectrum is a superposition
of local resonance. (c) The uniform precession mode of FMR can be
scattered to degenerate spin-wave mode. (d) In-plane FMR linewidth
for $f$=20 GHz as a function of field direction for FCCS samples
with different Co concentration; solid lines are fits to the data
using \ref{eq:Linewidth_all}. (e) Frequency dependence of FMR linewidth
for magnetization along {[}110{]} and {[}100{]} directions. The solid
lines are fits to the data using \ref{eq:Linewidth_all} and \ref{eq:TMSLinewidth}.\label{fig:IPR-linewidth}}
\end{figure*}

Using out-of-plane FMR measurements, we determined the intrinsic damping
by purposely suppressing the extrinsic contributions causing non-linear
frequency dependence \citep{NTU_FMR,lindnerTMSAngle_2009,okada2017magnetization}.
In practical applications, the relaxation mechanism for magnetization
in the film-plane is of primary importance. Hence, we investigated
the FCCS samples using in-plane magnetic field. The azimuthal angle
($\varphi_{H}$) dependence of resonance field shows four-fold symmetry
(\ref{fig:IPR-ku}(a)), indicative of a dominant in-plane cubic magneto-crystalline
anisotropy. Uniaxial anisotropy on the other hand, is negligible.
The in-plane magnetic easy axis is along {[}110{]} of the substrate
as the minimum of resonance field occurs at $\varphi_{H}=\pi/4$ and
$\varphi_{H}=3\pi/4$, along which the lattice constant of FCCS matches
MgO. We fit the data using the following equation derived from Smit-Beljers
formula \citep{SmitBeljers1955,Farle_review_1998}: 
\begin{equation}
2\pi f=\gamma\sqrt{UV}\label{eq:Kittel_angular}
\end{equation}
where $U=H\cdot\cos(\varphi_{M}-\varphi_{H}){\rm +}4\pi M_{{\rm eff}}+K^{4\parallel}/2M_{s}\cdot\left[3+\cos4(\varphi_{M}-\pi/4)\right]$
and ${\rm V=}H\cdot{\rm cos}(\varphi_{M}-\varphi_{H})+2K^{4\parallel}/M_{s}\cdot\cos4(\varphi_{M}-\pi/4)$,
$\varphi_{M}$ and $\varphi_{H}$ are the orientation of magnetization
and external field with respect to {[}100{]} direction of MgO. As
shown in \ref{fig:IPR-ku}b, the in plane four-fold anisotropy energy
increases with Co composition. The corresponding anisotropy field
$H^{4\parallel}=2K^{4\parallel}/M_{s}$ observed in Fe$_{2}$CoSi
is approximately 190$\,$Oe. The $\varphi_{H}$ dependence of FMR
linewidth changes dramatically with $x$, \ref{fig:IPR-linewidth}(d).
For Fe$_{2}$CrSi, we observe a nearly four-fold symmetry with minimum
at $\varphi_{H}$= 0 ({[}100{]}), conversely, smallest linewidth occurs
around $\varphi_{H}=\pi/4$, characteristic of Fe$_{2}$CoSi. Notably,
the linewidth anisotropy is much lower for FCCS films than for Fe$_{2}$CrSi
and Fe$_{2}$CoSi. 

The anisotropy in the linewidth may be attributed to angle dependent
damping\citep{1967_anisotropyDamping} or an extrinsic origin\citep{CFA_PRB_TMS_Cos}.
Recent calculations show that anisotropy in intrinsic damping is small
at room temperature\citep{AnisotropyDamping_seib,AnistropyDamping_Gilmore_theroy2010}.
Therefore, extrinsic linewidth broadening mechanisms, such as angle
dependent inhomogeneous contribution and spin wave scattering should
be considered. In general, the measured linewidth is the sum of intrinsic
and extrinsic contributions \citep{suhlAngular1955,1967_anisotropyDamping,Okada_linewidth}: 

\begin{equation}
\Delta H=\Delta H_{\mathscr{\textrm{int}}}+|\frac{\partial H_{\mbox{Res}}}{\partial4\pi M_{{\rm eff}}}|\Delta4\pi M_{{\rm eff}}+|\frac{\partial H_{\mbox{Res}}}{\partial H_{K}}|\Delta H_{K}+\Delta H_{TMS}\label{eq:Linewidth_all}
\end{equation}

\begin{figure}
\centering{}\includegraphics[width=0.85\columnwidth]{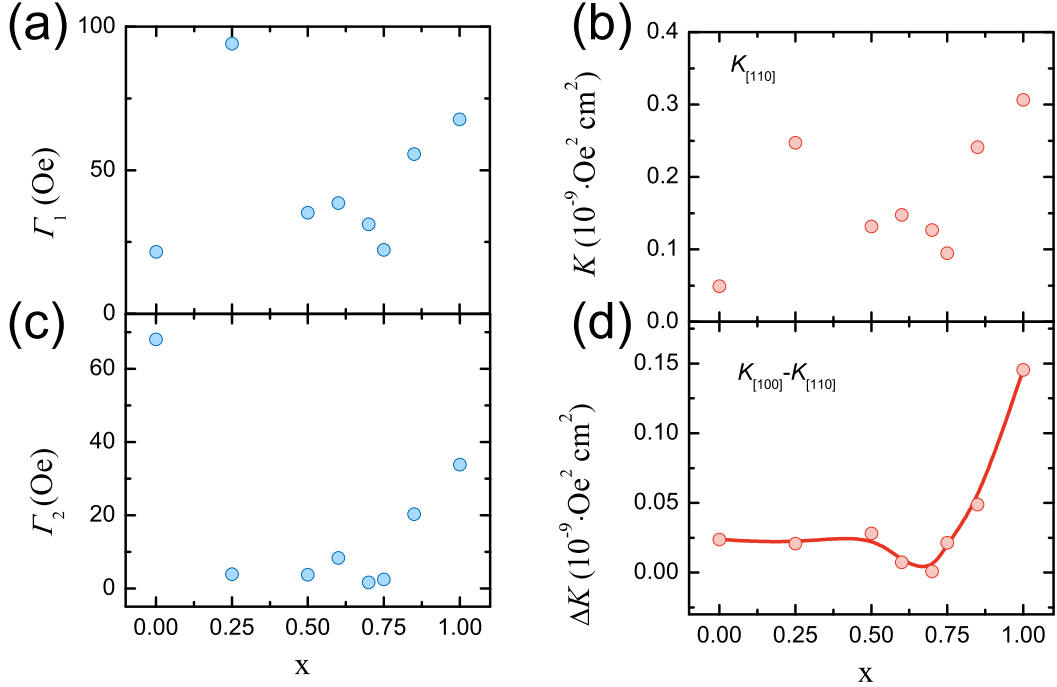}\protect\caption{Isotropic and anisotropic TMS contributions in Fe$_{2}$Cr$_{1-x}$Co$_{x}$Si.
(a, c) Isotropic and anisotropic components of the TMS linewidth determined
by the phenomenological method of angular dependent linewidth. (b,
d) Co concentration dependence of $K$ for magnetization along {[}110{]}
and anisotropy of $K$. \label{fig:FMR-TMS-results}}
\end{figure}

The first term on the right-hand side is the intrinsic linewidth (\ref{fig:IPR-linewidth}(a)
). For non-collinear magnetization and external field ($\varphi_{H}\neq\varphi_{M}$),
the magnetization direction changes during field sweep, giving additional
broadening to the intrinsic linewidth. In the present case of a relatively
small in-plane anisotropy, the effect is calculated numerically to
the first order using the resonance field together with the equilibrium
condition for magnetization\citep{suhlAngular1955,1967_anisotropyDamping,Platow_linewidth_all}:
$\Delta{\rm H}_{{\rm int}}=\frac{1}{{\rm |}2\pi\partial f/\partial H{\rm |}}\frac{\alpha}{\gamma}(U+V)\approx\frac{4\pi\alpha}{\gamma\cos(\varphi_{M}-\varphi_{H})}f$.
The second and third terms are due to dispersion of the effective
magnetization and anisotropy field, respectively (\ref{fig:IPR-linewidth}(b)).
The partial derivatives are evalued numerically at the resonance condition.
The last term accounts for two magnon scattering (TMS) \citep{Arias-PRB-1999,TMS_hurben}
activated by defects such as dislocations (See Supplemental Material
Fig. S4 for TEM results), representing relaxation of the uniform mode
to spin wave modes of non-zero wave vector (\ref{fig:IPR-linewidth}(c)).
We fit the data in \ref{fig:IPR-linewidth}d by adopting a phenomenological
form of TMS linewidth $\Delta H_{TMS}=\varGamma_{1}+\varGamma_{2}cos^{2}(2\varphi_{M}-\phi_{\textrm{max}})$
\citep{CFA_PRB_TMS_Cos,CFA_DOS}. Here, $\varGamma_{1}$ and $\varGamma_{2}$
are the isotropic and anisotropic TMS amplitude, respectively. The
anisotropic term correlates to defects of preferential orientation
\citep{TMS_ripple_2013}. $\phi_{\textrm{max}}$ is the orientation
along which the 2D Fourier transformation of defects potential shows
a maximum. We fit the data for all Co dopings. The extracted parameters
are shown in \ref{fig:FMR-TMS-results}(a) and (c). Although the anisotropic
extrinsic damping of FCCS appears universal in Hesuler alloys \citep{CoFeSi_anisotropy_2014},
when compared to Fe$_{2}$CrSi and Fe$_{2}$CoSi, is considerably
lower and almost vanishes for $0.5\leq x\leq0.75$. 

To study further the extrinsic relaxation and its anisotropy, we measured
the frequency dependence of FMR linewidth for magnetization along
the in-plane easy MgO{[}110{]} and hard MgO{[}100{]} axes. \ref{fig:IPR-linewidth}(e)
depicts significant differences between the two magnetization orientations
observed only for $x$ =0 and $x$ =1, in agreement with the angular
dependence data. On the other hand, a non-linearity as a signature
of TMS, was observed in all samples. We calculated the magnon wave
vector ( $k_{0}$) at which the spin wave energy degenerates with
the uniform modes $\omega(k_{0})=2\pi f$ for a given propagation
direction $\varphi$ with respect to magnetization in the film plane
\citep{TMS_ripple_2013,heinrich_TMS}. The total contribution to TMS
was then estimated using \citep{TMS_kink_fit_PRB,TMS_kink_model}
: 

\begin{equation}
\Delta H_{TMS}=\frac{2K}{\gamma^{2}}\frac{\partial f}{\partial H}\int\frac{k_{0}}{df/dk}d\varphi\label{eq:TMSLinewidth}
\end{equation}
 where $df/dk$ was evaluated at $k_{0}$ and $\partial f/\partial H$
calculated from the FMR resonance conditions. $K=0.16h^{2}A_{\xi}$
is the only fitting parameter related to defects. The fits to the
frequency dependence of linewidths are shown by solid lines. The fitting
parameter $K$ for both magnetization orientation and their difference
$\Delta K=K_{[100]}-K_{[110]}$ are indicative of the strength of
TMS and its anisotropy, respectively. The fitting results are plotted
in \ref{fig:FMR-TMS-results}(b) and (d). The Co doping dependence
of $K_{[110]}$ agrees with the trend obtained earlier from the angular
dependence of the linewidth. 

The anisotropy in TMS ($\Delta K$) is maximum for $x$=1 and negligible
for $x$ around 0.7. Furthermore, we obtain a relatively small anisotropy
of TMS for $x$=0, as expected from the weak non-linearity of the
two curves shown in the top panel of \ref{fig:IPR-linewidth}(e).
However, the results appear inconsistent with the phenomenological
fit shown in \ref{fig:FMR-TMS-results}(a). We attribute this to orientation
dependent long range inhomogeneity in the sample, apparent also from
the large zero frequency broadening ($\Delta H_{\textrm{0}}$) in
\ref{fig:FMR-Co-composition}(a) for $x$=0. Here, the phenomenological
fit cannot separate the angular dependence of inhomogeneous broadening
from four-fold TMS hence, overestimating the TMS contribution for
$x$=1. Nevertheless, both the frequency dependence and angular dependence
of the linewidths are in agreement, confirming that the isotropic
and anisotropic extrinsic damping constants of FCCS are significantly
reduced for $x$=0.5-0.75. Therefore, in FCCS with $0.5\leq x\leq0.75$,
low damping and nearly isotropic response are promising for practical
applications.

In summary, we studied the structure order and composition dependence
of magnetization relaxation in Fe$_{2}$Cr$_{1-x}$Co$_{x}$Si Heusler
alloy thin films. We show the ordered phase has a low intrinsic damping,
tunable by Co doping. Notably, the anisotropic extrinsic relaxation
of the quaternary alloys can be significantly reduced for FCCS films
with $x=$0.5\textasciitilde{}0.75. Furthermore, perpendicular and
in-plane magnetic anisotropy show nearly linear dependence on Co doping.
The present work demonstrates the tunable nature of both static and
dynamic magnetic properties of Fe-based Heusler alloys. The engineered
low damping constant in Fe$_{2}$Cr$_{1-x}$Co$_{x}$Si thin films
encourages the utilization of Heusler alloys in room temperature spintronics. 
\begin{acknowledgments}
The work was supported by the Ministry of Education (MOE), Academic
Research Fund (AcRF) Tier 2 Grant (MOE2014-T2-1-050), The A{*}STAR
Pharos Fund (1527400026), and the National Research Foundation (NRF),
NRF-Investigatorship (NRF-NRFI2015-04). The authors acknowledge the
Singapore Synchrotron Light Source (SSLS) for providing the facilities
to perform x-ray experiments. The Laboratory is a National Research
Infrastructure under the National Research Foundation Singapore.
\end{acknowledgments}

\bibliographystyle{apsrev4-1}

\end{document}